\documentstyle[12pt]{article} \long\def\del#1\enddel{ }
\let\3=\ss \catcode`\"=\active \let"=\"
  \oddsidemargin=\evensidemargin
\let\ni=\noindent  
 \let\msk=\medskip \let\bsk=\bigskip
  
\let\qd=\quad   \def\ve{\vfil\eject}

\let\a=\alpha    \let\e=\varepsilon
\let\z=\zeta  \let\th=\theta  
\let\l=\lambda \let\m=\mu   \let\p=\pi \let\r=\rho
   \let\c=\chi
 
\let\Ph=\Phi  \let\Ps=\Psi   
   
\def\0{\over }    \def\1{\vec }   \def\2{{1\over2}} \def\3{{\ss}}
\def\4{{1\over4}} \def\5{\overline }   \def\6{\partial } \def\7#1{{#1}\llap{/}}
\def\8#1{{\textstyle{#1}}}        \def\9#1{{\bf {#1}}}
\def\_#1{$\underline{\hbox{#1}}$} \def\^#1{$\overline{\hbox{#1}}$}

\def\<{\langle } \def\>{\rangle }  
 
\def \({\left( } \def \){\right) }

\let\ap=\approx     \let\aus=\in
      \let\and=\wedge
     
\def\|#1{{}_{\bigg|_{#1}}}

\def\mao#1{\mathop{\rm {#1}}\nolimits}    \def\tr{\mao{tr}} 
  
 \def\gcd{\mao{gcd}} 
\ifx\Box  \else \fi

\def\pmbf#1{\setbox0=\hbox{${#1}$}   \kern-.025em\copy0\kern-\wd0
      \kern.05em\copy0\kern-\wd0     \kern-.025em\raise.0433em\box0 }

 \def\cm{{\cal M}} \def\co{{\cal O}} 
   
\def\inbar{\vrule height1.5ex width.4pt depth0pt} 
\def\ifundefined#1{\expandafter\ifx\csname#1\endcsname\relax}
\makeatletter \ifundefined{new@mathgroup} {} \else 
\del
\new@mathgroup\sffam
\new@internalmathalphabet\mathsf\sffam{cmss}{m}{n}
\def\psf{\fontfamily\sfdefault \fontseries\default@series
    \fontshape\default@shape\selectfont\mathsf}
\enddel
\fi
\def\ZZ{\relax{\sf Z\kern-.4em \sf Z}}  \def\IR{\relax{\rm I\kern-.18em R}}
\def\IN{\relax{\rm I\kern-.18em N}} \def\IP{\relax{\rm I\kern-.18em P}}
\def\IQ{\relax\,\hbox{$\inbar\kern-.3em{\rm Q}$}}
\def\IC{\hbox{\,$\inbar\kern-.3em{\rm C}$}}
\def\citen#1{\if@filesw \immediate\write \@auxout {\string\citation{#1}}\fi%
\@tempcntb\m@ne \let\@h@ld\relax \def\@citea{}%
\@for \@citeb:=#1\do {\@ifundefined {b@\@citeb}%
    {\@h@ld\@citea\@tempcntb\m@ne{\bf ?}%
    \@warning {Citation `\@citeb ' on page \thepage \space undefined}}%
    {\@tempcnta\@tempcntb \advance\@tempcnta\@ne
    \setbox\z@\hbox\bgroup\ifcat0\csname b@\@citeb \endcsname \relax
       \egroup \@tempcntb\number\csname b@\@citeb \endcsname \relax
       \else \egroup \@tempcntb\m@ne \fi \ifnum\@tempcnta=\@tempcntb
       \ifx\@h@ld\relax \edef \@h@ld{\@citea\csname b@\@citeb\endcsname}%
       \else \edef\@h@ld{\hbox{--}\penalty\@highpenalty
	      \csname b@\@citeb\endcsname}\fi
    \else \@h@ld\@citea\csname b@\@citeb \endcsname \let\@h@ld\relax \fi}%
 \def\@citea{,\penalty\@highpenalty\hskip.13em plus.13em minus.13em}}\@h@ld}
\def\@citex[#1]#2{\@cite{\citen{#2}}{#1}}%
\def\@cite#1#2{\leavevmode\unskip
  \ifnum\lastpenalty=\z@\penalty\@highpenalty\fi
  \ [{\multiply\@highpenalty 3 #1
  \if@tempswa,\penalty\@highpenalty\ #2\fi}]}   
\makeatother 

\def\beq{\begin{equation}} \def\eeq{\end{equation}} \def\eql#1{\label{#1}\eeq}
\def\bea{\begin{eqnarray}} \def\eea{\end{eqnarray}} 
\def\fnote#1#2{\begingroup\def\thefootnote{#1}\footnote{#2}
	   \addtocounter{footnote}{-1}\endgroup}

\def\plb#1 #2 {Phys. Lett. {\bf B#1} #2 }
\def\phr#1 #2 {Phys. Rep. {\bf  #1} #2 } 
\def\npb#1 #2 {Nucl. Phys. {\bf B#1} #2 }
\def\aph#1 #2 {Ann. Phys. {\bf #1} #2 }  \let\ap=\aph
\def\jmp#1 #2 {J. Math. Phys. {\bf #1} #2 }
\def\prd#1 #2 {Phys. Rev. {\bf D#1} #2 }
\def\prl#1 #2 {Phys. Rev. Lett. {\bf #1} #2 }
\def\rmp#1 #2 {Rev. Mod. Phys.  {\bf #1} #2 }
\def\zpc#1 #2 {Z. Phys. {\bf #1C} #2 }
\def\cmp#1 #2 {Comm. Math. Phys. {\bf #1} #2 }
\def\mpl#1 #2 {Mod. Phys. Lett. {\bf A#1} #2 }
\def\ijmp#1 #2 {Int. J. Mod. Phys. {\bf A#1} #2 }

\def\naive{na\"\i ve} \def\[{\left[} \def\]{\right]} \let\cdots=\ldots
\def\ng{p_{11}} \def\na{p_{12}} 
\def\tbf#1:{{\noindent\bf #1:}}            \long\def\new#1\endnew{{\bf #1}}
   
\def\PP{Poincar\'e polynomial}  \def\PD{Poincar\'e duality}
\def\LG{Landau--Ginzburg}   \def\LGO{\LG\ orbi\-fold}

\let\Ph=X  \let\Ps=Y

\def\figuresonly{\pagestyle{empty}\figa\ve\figb\ve\figc\end{document}}
\long\def\old#1\endold{{\small #1}}         \def\oldansw{o } \def\cutansw{c }
\newcount{\npre} \npre=1  \def\negansw{s } \def\figansw{f } \def\textansw{t }
\def\ifpre{\ifnum\npre=1 } \def\ifsub{\ifnum\npre=0 }        \def\cut#1{#1}
\def\askversion{\message{
Preprint (p) / submit (s) / text only (t) / figures only (f):  (p/s/t/f)? }
    \read-1 to\answ \ifx\answ\negansw \npre=0 \else \npre=1 \fi
    \ifx\answ\figansw { } \else \def\figuresonly{ }   \fi
    \ifx\answ\oldansw \def\old##1\endold{{\small ##1}}\fi
    \ifx\answ\textansw \npre=2 \else \message{
Cut figures (use 'c' in case of memory problem):  (c/n)? }
    \read-1 to\answ\ifx\answ\cutansw \def\cut##1{}\npre=7\fi\fi \figuresonly }

\def\bpic{\begin{picture}} \def\epic{\end{picture}} \thicklines
\def\lab#1)#2#3{\put#1){\makebox(0,0)[#2]{\small #3}}}
\def\putlin#1,#2,#3,#4,#5){\put#1,#2){\line(#3,#4){#5}}} 
\def\putvec#1,#2,#3,#4,#5){\put#1,#2){\vector(#3,#4){#5}}}

\newcount{\vmul} \newcount{\hdiv} 
\newcount{\wi}   \newcount{\he}   
\newcount{\hq}   \newcount{\vq}   
\newcount{\hoff} \newcount{\auxc} \newcounter{figco}   \def\npt{\circle*{2}}

\def\vlline{\put(-3,0){\line(1,0)6}}      
\def\vlright#1{\put(6,0){\makebox(0,0)[l]{\scriptsize #1}}}
\def\putvl#1{\mbox{\bpic(0,0)\funit=1pt\vlline\vlright{#1}\epic}}
\def\putvm#1{\mbox{\bpic(0,0)\funit=1pt\vlline\epic}}   
\def\vlab#1{\vq=#1\multiply\vq by\vmul \put(-\hoff,\vq){\putvl{#1}}
	    \put(\hoff,\vq){\putvm{#1}} }

\def\hlline{\put(0,-3){\line(0,1)6}}     
\def\hltop#1{\put(0,6){\makebox(0,0)[b]{\scriptsize #1}}}
\def\puthl#1{\mbox{\bpic(0,0)\funit=1pt\hlline\hltop{#1}\epic}}
\def\hlab#1{\hq=#1\divide\hq by\hdiv \put(\hq,0){\puthl{#1}} }    
\def\hlabo{\put(0,0){\mbox{\bpic(0,0)\funit=1pt\put(0,-3){\line(0,1)3}\epic}}}

\def\Vpt#1,#2){\hq=#2\advance\hq by -#1 \multiply\hq by 2 \divide\hq by\hdiv
	       \vq=#1\advance\vq by #2 \multiply\vq by\vmul\put(\hq,\vq){\npt}}
\def\Vplo#1{\vbox{\hdiv=2\vmul=1 \figsca \auxc=\he \multiply\auxc by\vmul
    \hoff=\wi\divide\hoff by2 \stepcounter{figco}\message{[Fig. \arabic{figco}}
    \begin{center}\let\.=\Vpt \bpic(\wi,\auxc)(-\hoff,0) \figlab #1 \hlabo
    \put(-\hoff,0){\framebox(\wi,\auxc){}} \epic \\[5mm]
    Fig. \arabic{figco}: \figcap \end{center}} \vfil \message{]}}
\def\figsca{\unitlength=1.1pt \wi=500 \he=400} \let\funit=\unitlength

\begin{document}
\def\hannover{ITP--UH--16/94} \def\wien{TUW--94/20}
{\hfill hep-th/9412033   \vskip-9pt \hfill\wien\vskip-9pt  \hfill \hannover}
\vskip 15mm \centerline{\hss\Large\bf
       Landau--Ginzburg orbifolds with discrete torsion \hss}
\begin{center} \vskip 8mm
       Maximilian KREUZER\fnote{*}{e-mail: kreuzer@tph.tuwien.ac.at}
\vskip 3mm
       Institut f"ur Theoretische Physik, Technische Universit"at Wien\\
       Wiedner Hauptstra\3e 8--10, A-1040 Wien, AUSTRIA
\vskip 6mm               and
\vskip 3mm
       Harald SKARKE\fnote{\#}{e-mail: skarke@kastor.itp.uni-hannover.de}
\vskip 3mm
       Institut f"ur Theoretische Physik, Universit"at Hannover\\
       Appelstra\3e 2, D--30167 Hannover, GERMANY

\vfil                        {\bf ABSTRACT}                \end{center}

We complete the classification of (2,2) vacua that can be constructed from
Landau--Ginzburg models by abelian twists with arbitrary discrete torsions.
Compared to the case without torsion the number of new spectra is
surprisingly small. In contrast to a popular expectation mirror
symmetry does not seem to be related to discrete torsion (at least not in
the present compactification framework): The Berglund--H"ubsch construction
naturally extends to orbifolds with torsion; for more general potentials, on
the other hand, the new spectra neither have nor provide mirror partners
in our class of models.

\vfil\noindent \hannover\\[1pt] \wien\\[3pt] November 1994 \msk
\thispagestyle{empty} \newpage
\setcounter{page}{1} \pagestyle{plain}
\ifsub \baselineskip=20pt \else \baselineskip=14pt \fi

\section{Introduction}

Much of the early string phenomenology was based on a few examples of
4-dimensional string constructions. Only more recently there was some
effort to make theoretical and phenomenological statements on the basis of
explicit classifications.
In a pioneering work Schellekens and Yankielowicz \cite{sy,fkss} generated huge
lists of string vacua that can be constructed with non-diagonal modular
invariants of tensor products of minimal models. Although their approach was
to perform a statistical search, their results are fairly complete at least
for (2,2) models, and probably also for the (2,1) case.
The cases of free fermions and free bosons, which appear to be quite attractive
from a phenomenological point of view~\cite{z2z2,fiqs}, seem to be far less
tractable as far as classification is concerned
and the investigations were mainly limited
to single twists \cite{abel,orbi}.

The constructive classification can be carried through very far for \LGO s,
which contain Gepner's models as a small subclass.
The authors of ref.~\cite{cls} already constructed a large number
of Calabi--Yau hypersurfaces in weighted projective space.
Although no strict limits on the degrees and on the singularity types were
known at that time, the majority of all models in this class was found.
This construction was completed and extended to
Landau--Ginzburg models \cite{lvw} in refs. \cite{nms,klesch}.

In \cite{aas} we computed all abelian orbifolds \cite{v,iv}
of such models without any restrictions on the twists except for the (2,2)
condition.
In that work, however, we discarded
the possibility of introducing non-trivial phases, the so-called discrete
torsions \cite{dt}, among the generators of the twist group.
When one tries to include such phases, two major problems occur:
First, the (\naive) number of possibilities is so enormous in some cases
that the redundancies due to permutation symmetries of the factors in a
tensor product have
to be eliminated or at least reduced. Second, we need 
some new efficient algorithms for the computation of spectra
in the presence of torsion.
This problem arises for general
singularity structures, where an explicit basis of the chiral ring would have
to be constructed on a case by case basis, so that the projection approach
is hardly managable (see below).

Fortunately the two problems do not occur simultaneously: Large permutation
symmetries arise mainly from Landau--Ginzburg potentials with a simple
singularity structure.
Therefore we first considered the case of ADE models, i.e. tensor
products of minimal models, for which we completed the classification of (2,2)
models in ref. \cite{ade}. In that class all diagonal abelian
symmetries happen to be generated by simple currents. So the orbifolds
with torsions coincide with the $N=2$ minimal models
with simple current modular invariants~\cite{sci} (the exceptional
invariant $E_7$ has to be taken into account explicitly in both frameworks).
Indeed, we reproduced all spectra of ref. \cite{sy} and found the 4 missing
ones.%
\footnote{Schellekens then found the additional spectra by changing the
	  statistics in their original program.}

In the present note we complete the enumeration of abelian (2,2)
Landau--Ginzburg orbi\-folds.
The results for the spectra are not too interesting from a
practical point of view, since the eventual breaking of $E_6$ will in general
change the numbers of generations~\cite{fiqs}.
Nevertheless, most of the effort in the literature has been focused on the
(2,2) case, which allows for general comparisons of constructions and
investigations on theoretical subjects like mirror symmetry
\cite{gp,cls,bh,mmi}.

In section~2 we review the construction of \LGO s and derive constraints on
the general structure of spectra that we have to expect. These allow us to
reconstruct the complete Hodge diamond from a few quantities that can be
calculated efficiently for an arbitrary singularity type.
In section~3 we describe the algorithms that we use for our computations
and our strategies to avoid redundancies.
At last we present  our results and discuss some implications and prospects
of further investigations.

\section{
\LGO s and spectra}

The \LG\ description of (super) conformal field theories relies on the
assumption that, for a given lagrangian, there exists a fixed point of the
renormalization group flow at which the quantum field theory is conformally
invariant.
In case of $N=2$ supersymmetry the superpotential is expected not to be
renormalized. Therefore, for a quasi-homogeneous superpotential $W(X_i)$, the
chiral ring turns out to be isomorphic to the local algebra of the
singularity $\IC[X_i]/\<\6_iW\>$, i.e. the quotient of the polynomial ring by
the ideal generated by the gradients (which become descendant fields via their
equations of motion).
For a string vacuum with space-time supersymmetry we project to integral
$U(1)$ charges, and we need to include the twisted sectors in order
to respect modular invariance.
It has been shown by Vafa~\cite{v} that the invariant states have integral
charges in all sectors if the central charge $c=3D$ is a multiple of 3.

Of course we can consider more general orbifolds. For a (2,0) model the
projector $j=\exp(2\p i J_0)$ to integral left charges has to be contained in
the twist group and the left-moving spectral flow operator, which arises as the
ground state in the sector twisted by $j^{-1}$, should be invariant.
In ref. \cite{iv} the left and right charges
\beq
     Q_\pm|h\>=\sum_{\th_i^h\not\aus\ZZ}~\((\8\2-q_i)\pm(\th_i^h-\8\2)\)|h\>
\eql{charge}
and the transformation properties under group elements $g$
\beq
     g|h\>=(-1)^{K_gK_h}\e(g,h){\det g_{|_h}\0 \det g}|h\>
\eql{phase}
of the ground state $|h\>$ in the sector twisted by $h$ were derived.
In these formulas $Q_\pm$ are the eigenvalues of $J_0$ and $\5J_0$.
The commuting group elements $g$ and $h$ are assumed to act diagonally.
$h$ acts on the superfield $X_i$ by multiplication with a phase factor
$\exp(2\p i\th_i^h)$, and $\det g_{|_h}$ is the determinant of the
representation of $g$ restricted to the fields that are invariant under $h$.
The discrete torsions%
\footnote{As usual, $\e(g,h)$ have to be multiplicative in both entries
	  and must fulfill $\e(g,h)\e(h,g)=\e(1,g)=1$.}
$\e(g,h)$ and the signs $(-)^{K_g}$
parametrize the phase choices of the action of $g$ in the sector $h$ that are
allowed by modular invariance; note that the group actions in the Ramond
sector are fixed by those in the NS sector only up to the signs $(-)^{K_g}$
which we are, a priory, free to choose.
The formulas (\ref{charge},\ref{phase}) refer to the NS sector;
the corresponding results in the Ramond sector are obtained by spectral flow.

For a (2,0) vacuum we require
\beq 	\e(j,g)=(-1)^{K_gK_j}\det g		\eql{21-con}
to ensure that $|j^{-1}\>$ (corresponding to the
holomorphic three--form in the Calabi--Yau context) is invariant~\cite{iv}.
In fact, such vacua 
have (2,1) supersymmetry since the alignment of R and NS states
in tensor products is implicit in the \LG\ framework:
In the sum over spin structures the fermionic components of all
superfields $X_i$ are understood to have identical periodicity properties;
(2,0) vacua could, however, be obtained by an explicit tensoring of building
blocks that are described by \LGO s. Since we 
need the left-moving
supersymmetry, modular invariance strongly constrains such a construction.

For (2,2) supersymmetry we also need the right--handed analogue of
(\ref{21-con}), which
implies $(-1)^{K_g}=\det g$, so that all twists must have real
determinants $\pm 1$.
Then the group transformations take the form
\beq  g|h\>=(-1)^{K_g(K_h-1)}\e(g,h)\det g_{|_h}|h\>       \eql{phase2}
and, since $\det j=\exp(2\p i\sum q_i)$ and $D=\sum 1-2q_i$,
\beq \e(j,g)=(-1)^{K_g(N-D+1)},           \eql{ejg}
where $N$ is the number of basic superfields $X_1,\ldots,X_N$.

Let us denote by $p_{ij}$ the number of states with left and right $U(1)$
charges $(q_L,q_R)=(i,j)$. The untwisted sector only
contributes to states with $q_L=q_R$.
For (2,2) vacua there is the \PD\ $p_{i,j}=p_{D-i,D-j}$:
Using spectral flow, this can be understood as a
consequence of charge conjugation in the Ramond sector~\cite{lvw}.
The explicit mapping of states takes the following form:
For any monomial $M(X_i)$ there exists a monomial $\5M(X_i)$ such that
$M\,\5M$ is the unique monomial of highest weight. As that monomial is
invariant under transformations with ~$\det=\pm1$, $M$ and $\5M$ must
transform with complex conjugate phases under any such linear symmetry.
This duality is now easily extended to the full orbifold:
Observe that the
action of $g$ on $|h\>$ and the action of $g^{-1}$ on $|h^{-1}\>$, as given by
eq.~(\ref{phase2}), are the same up to a factor ~$(\det g_{|_h})^2$.
This factor is exactly cancelled by the phase difference
between the actions of $g$ on $M_h$ and of $g^{-1}$ on $\5M_h$, where $M_h$
and $\5M_h$ are monomials of $h$-invariant fields such that their product is
the highest weight monomial of the $h$-invariant \LG\ model.
Furthermore, the sums of the charges of $|h\>$ and $|h^{-1}\>$ are
$Q_\pm=(c-c_h)/3$, where $c_h$ is the contribution of the untwisted fields
to the central charge.
Hence, if $M_h|h\>$ contributes to $p_{ij}$ then $\5M_h|h^{-1}\>$ contributes
to $p_{D-i,D-j}$.

We get further restrictions if $i=0$ or $j=0$. Consider, for example,
states with vanishing left charge $Q_+=0$. We have shown in~\cite{nms}
that such a state must be a twisted ground state $|j_a\>$, where the group
element $j_a$ vanishes on a subset
of the fields and acts like $j$ on the remaining fields.
Furthermore, the contribution of the invariant fields
to the central charge
must be $3(D-Q_-)$, where $Q_-$ is the right charge of $|j_a\>$.          %
It is easy to see that invariance of $|j_a\>$ under all elements of the
centralizer of $j_a$ implies invariance of $|j/j_a\>$.
Hence $p_{0j}=p_{0,D-j}$, and by a similar reasoning $p_{i,0}=p_{D-i,0}$.
For $D=3$ the same result can be obtained for a general (2,2) model from
the charge sum rule $\tr_R (-)^FJ_0^2=(D/12)\tr_R(-)^F$ that was derived
in~\cite{sum}. Equality of
\beq     \2\tr(-)^F=p_{20}-p_{10}+p_{02}-p_{01}+p_{11}-p_{21}   \eeq
and
\beq  2\tr(-)^F(J_0)^2=p_{20}-p_{10}+9(p_{02}-p_{01})+p_{11}-p_{21} \eeq
implies $p_{01}=p_{02}$; the same consideration for the right moving
charges implies $p_{10}=p_{20}$.

For $p_{01}$ we can also derive the upper limit $D$.
To arrive at this result we show that if the two twists $j_a$ and $j_b$
contribute to $p_{01}$ this implies that $j_a$ and $j_b$ cannot both act on the
same non-trivial field (or must be identical): Let $\cm$ be the set of fields
on which both $j_a$ and $j_b$ act with a non-trivial phase, i.e. $j_aX_i=
j_bX_i=jX_i$ for $X_i\in\cm$.
Invariance of $|j_b\>$ and $|j_a\>$ under all twists implies
\beq j_a|j_b\>=(-1)^{K_{j_a}K_{j_b}}\e(j_a,j_b)\(\det j_a{|_\cm}\)^{-1}|j_b\>
              =|j_b\> \eeq
and the analogous equation with $j_a$ and $j_b$ exchanged. From
$\det j_a{|_\cm}=\det j_b{|_\cm}$ we conclude that $\e(j_a,j_b)=\pm1$ and
thus $\det j_a{|_\cm}=\pm1$. Finally, there are only three non-degenerate
configurations with $c=3$, namely $\IC_{(111)}[3]$, $\IC_{(12)}[6]$ and
$\IC_{(11)}[4]$, and the determinants of the actions restricted to $\cm$ can
only be real if $\cm$ is empty or if $j_a=j_b$. This proves the above
statement.
As a corollary we can also show that $p_{01}$ can never be $D-1$:
If the states $|j_a\>$ with $1\le a\le D-1$ contribute to $p_{01}$,
then $|j_D\>=|j/(j_1\cdots j_{D-1})\>$ also contributes.

Let us sum up our results on the form of the ``Hodge diamond'' for $D=3$.
\PD\ $p_{i,j}=p_{D-i,D-j}$ and the results $p_{0,i}=p_{0,D-i}$ and
$p_{i,0}=p_{D-i,0}$ derived above imply $p_{11}=p_{22}$, $p_{12}=p_{21}$ and
$p_{33}=p_{30}=p_{03}=p_{00}=1$.
If one of the numbers $p_{01}=p_{02}=p_{31}=p_{32}$ or
$p_{10}=p_{20}=p_{13}=p_{23}$, which can only assume the values 0, 1 or 3,
is different from zero, either gauge symmetry or
space-time supersymmetry is enhanced.
In this case the possibility of having chiral fermion generations is
excluded, implying $p_{11}=p_{12}$.
\del
These numbers
 From CFT we know that in the context of the
heterotic string a positive $p_{01}$ or $p_{10}$ implies an extension of the
gauge group to (at least) $E_7$ or extended space-time supersymmetry.
As this excludes the possibility of having chiral fermion generations,
$p_{01}$ or $p_{10}$ can be non-zero only if the
Euler number $\c=2(p_{12}-p_{11})$ vanishes.

In fact, if $p_{01}=3$ or $p_{10}=3$ we always
find that $p_{ij}$ factorizes into left and right contributions, i.e. is equal
to $p_{i0}p_{0j}$; otherwise this product only gives a lower bound for
$p_{ij}$.
\enddel

In ref. \cite{iv} the possibility of twisting by group actions with negative
determinant in case of odd order $d$ of $j$ was excluded because of the
constraints on the discrete torsions with $j$:
Consider the case of odd $D$ and $d$, implying that $N$ is also
odd, $N=2n-1$.
Taking the $d^{\rm th}$
power of equation (\ref{ejg}), we see that we are actually restricted to
$\det g=1$.
This seems strange, however, because by adding a trivial variable $X_{2n}$
we can make $j$ even and thus get rid of this restriction.
In fact, even without resorting to trivial fields, it is easy
to see that, for odd $D$, the fields in the Ramond sector are quantized in
units of $1/2d$ rather that in units of $1/d$. Thus $j$ is of
order $2d$ rather than of order $d$, and a negative determinant need not
be excluded a priory if we double the range for twists by powers of $j$.
We do, however, agree with the conclusion of \cite{iv} in the following
sense: Modding by such symmetries can be disregarded, because the
resulting orbifolds cannot yield anything new. This can be seen explicitly
(at least at the level of particle spectra) from the following calculation:
After adding a trivial field $X_{2n}$,
the new $j$ decomposes as $j_{\rm odd}j_2$, where $j_{\rm odd}$ is
our original $j$ and $j_2$ acts only on $X_{2n}$. Then we have
\bea  g|hj_2\>&=&(-1)^{K_g(K_{hj_2}-1)}\e(g,hj_2)\det g_{|_{hj_2}}|hj_2\>\\
   &=& (-1)^{K_gK_h}\e(g,h)\det g_{|_{hj_2}}|hj_2\>                  \eea
and
\beq j_2|h\>=(-1)^{K_h-1}\det {j_2}_{|_h}|h\> .\eeq
We first consider groups generated by $j_2$ and by elements $g$ with
$\det g=1$ which do not act on $X_{2n}$. Such a $g$ acts on
$|hj_2\>$ in exactly the same way as it acts on $|h\>$, whereas the action
of $j_2$ on any state is trivial. This means that effectively all states
have been doubled by the introduction of $X_{2n}$. Let us examine now what
happens if we allow arbitrary group actions with $\det=\pm 1$.
It is easy to see that we can choose the group to be generated by elements
$g$ as above, by $j_2$ and by some generator $s$ with $\det s =1$
and $sX_{2n}=-X_{2n}$, i.e. $\det s_{|_{\{X_1,...X_{2n-1}\}}}=-1$. Then
\beq s|hj_2\>=\e(s,h)\det s_{|_{hj_2}}|hj_2\> =-\e(s,h)\det s_{|_h}|hj_2\>,
                                                                        \eeq
i.e. the phase with which $s$ acts on an $hj_2$-twisted state is minus the
phase with which it acts on the corresponding $h$-twisted state.
If an $h$-twisted state survives projections by all generators with
$\det{|_{\{X_1,...X_{2n-1}\}}}=1$ (in particular with $s^2$), then the
action of $s$ on this state must be $\pm1$ and exactly one
of the two states ($h$- or $hj_2$-twisted) will survive the $s$ projection.
Now let $h$ be a group element with $\det h_{|_{\{X_1,...X_{2n-1}\}}}=1$.
Then
\beq j_2|sh\>=(-1)^{K_s+K_h-1}\det {j_2}_{|_{sh}}|sh\> =(-1)^{K_h-1}(-1)^{K_h}
    =-|sh\>,                     \eeq
implying that no $sh$-twisted state can survive the $j_2$ projection.
Therefore the states of this model are in one to one correspondence with the
states of the model without $X_{2n}$, with the group actions restricted
to those with unit determinant.

\section{Computing spectra and reducing redundancies}

For models with a simple structure of the chiral ring (e.g., the case of ADE
models \cite{ade}) it is quite straightforward to
implement formulas (\ref{charge},\ref{phase}) in a computer program.
An alternative is provided by the formulas \cite{iv} (for a discussion
adequate to the present context, see \cite{aas})
\beq \5\c={1\0|G|}\sum_{gh=hg}(-1)^{N_h+K_gK_h+K_{g}}\e(g,h)
     \prod_{\th_i^g=\th_i^h=0} {n_i-d\0n_i},                           \eql{xb}
where $N_h$ denotes the number of $X_i$ invariant under $h$, and
\beq -\c={1\0|G|}\sum_{gh=hg}(-1)^{N+K_gK_h+K_{gh}}\e(g,h)
     \prod_{\th_i^g=\th_i^h=0} {n_i-d\0n_i}                           \eql x
for the dimension of the chiral ring and Witten's index, respectively.
For $D=3$ these numbers, together with $p_{10}$ and $p_{01}$, which are zero
for $\c\ne 0$, contain the full information about the spectrum.

In fact, a calculation of the \PP\ which does not
require an explicit basis of the chiral ring can be performed for arbitrary
\LGO s: The starting point is, of course, the formula
\beq
     P(t,\5t)=\prod_{i=1}^N{1-(t\5t)^{1-q_i}\01-(t\5t)^{q_i}}
\eql{pp}
for the \PP\ of an untwisted \LG\ model.
The inverse of the denominator of this expression is the \PP\ of the freely
generated polynomial ring $\IC[X_1,\ldots,X_N]$, and the factors in the
numerator 
correspond to dividing by
the ideal that is generated by the independent polynomials $\6_i W$.
In a diagonal basis the action of a group element $g$ of order $\co(g)$
is given by $gX_i=\r_i X_i$ with $\r_i=\z^{r_i}$, where $\z$
is an $\co(g)^{\rm th}$ root of unity. 
We can define a ``$g$-extended'' \PP\
\beq P(t,\5t;\tilde\z):=\sum_s\sum_{r=0}^{\co(g)-1}\m(r,s)\tilde\z^r(t\5t)^s,
                                                                    \eql{ppg}
where $\m(r,s)$ is the number of basis monomials of weight $s$ in the chiral
ring that transform with a phase $\z^r$ under the action of $g$, and
$\tilde\z$ is a formal variable subject to $\tilde\z^{\co(g)}=1$.
Then the standard arguments that lead to formula (\ref{pp}) give
\beq
     P(t,\5t;\tilde\z)=
\prod_{i=1}^N{1-\tilde\r_i^{-1}(t\5t)^{1-q_i}\01-\tilde\r_i(t\5t)^{q_i}}
\eql{phasex}
with $\tilde\r_i=\tilde\z^{r_i}$ whenever $W$ is invariant under the
action of $g$.
We can easily get rid of $\tilde\z$ in the numerator because of
\beq {1\0 1-\tilde\z^r x}=
   {1+\tilde\z^r x+ \ldots+(\tilde\z^r x)^{\co(g)-1}\0 1-x^{\co(g)}}, \eeq
where we used $\tilde\z^{\co(g)}=1$.
With this identity, 
(\ref{phasex}) can be recast into
\beq P(t,\5t;\tilde\z)=\sum_{r=0}^{\co(g)-1}\tilde\z^rP_r(t\5t) \eeq
with rational functions $P_r$. Of course, non-degeneracy and $g$-invariance
of $W$ imply that the $P_r$ must be polynomials.
\del 
the expression
a polynomial in $t\5t$ and $\tilde\z$ , divided by a polynomial in $t\5t$.
After taking the numerator modulo $\tilde\z^{\co(g)}-1$, the division can
be carried out
and, whenever $g$ is a symmetry of a non-degenerate polynomial $W$
that is quasi-homogeneous of degree 1 with respect to the given weights,
we must get a polynomial in $t\5t$ and $\tilde\z$. The coefficient
of $\tilde\z^i$, with $\tilde\z$ a formal variable, tells us the number of
states that
transform with a phase $\z^i$, with the specific root of unity inserted for
$\tilde\z$, as well as their charges.
\enddel
Now the projection to invariant states is trivial
(in the $h$-twisted sector we get an additional factor
$t^{Q_+^{(h)}}\5t^{Q_+^{(h)}}\tilde\z^{n{(h)}}$
describing the charges and transformation property of the twisted vacuum, and
the product in (\ref{phasex}) is restricted to fields invariant under $h$).
This algorithm, and its obvious extension to multiple projections, can be
implemented easily in an algebraic computer program. For our purposes,
however, it is too
slow since it would involve extensive polynomial algebra operations.

Returning to $D=3$ and integral charges, we better use (\ref{xb},\ref x)
and avoid the use of complex numbers, which come from the discrete torsions
$\e(g,h)$, in the following way:
Defining
\beq \a_h(g)=(-1)^{K_g(K_h+1)}\e(g,h),                  \eeq
we see that the $h$ twisted sector yields
\beq n_h={1\0|G|}(-1)^{N_h}\sum_{gh=hg}\a_h(g) \prod_{\th_i^g=\th_i^h=0}
                        {n_i-d\0n_i}                           \eql{nh}
states, which contribute with the sign $(-1)^{K_h+N-N_h}$ to the index
(\ref x), i.e. they increase even or odd Betti numbers depending on whether
$K_h+N-N_h$ is even or odd.
We can define
an equivalence relation among the elements of $G$ by $g\sim h$
if and only if $h=g^\l$ with $\gcd(\l,\co(g))=1$. Whether an $X_i$ is
invariant under
$g$ only depends on the corresponding equivalence class $[g]$ of $g$, not
on the choice of $g$ within such a class.
If $\co(g)$ is a power of a prime number $p$, then $[g]$ is the set of
all $g^\l$ for which $\l$ is not divisible by $p$.
We notice that $\a_h(g_1g_2)=\a_h(g_1)\a_h(g_2)$, $|\a_h(g)|=1$ and
$\a_h(1_G)=1$, i.e. $\a_h$ is a homomorphism from $G$ into $U(1)$.
Therefore $\a_h(g)$ is always a power of the $\co(g)^{\rm th}$ root
of unity and
\bea \a_h(g)+\a_h(g^2)+\ldots+\a_h(1)=&\co(g)\qd &\hbox{if}\qd\a_h(g)=1\\
                                      &0\qd &\hbox{otherwise}.      \eea
For $\co(g)=p^k$, where $p$ is a prime number, this implies
\bea \sum_{g\in[g]}\a_h(g)=& 0\qd &\hbox{if}\qd\a_h(g^p)\ne 1\\
                       & -p^{k-1}\qd &\hbox{if}\qd\a_h(g)\ne 1,\a_h(g^p)= 1\\
                       & p^k-p^{k-1}\qd&\hbox{if}\qd\a_h(g)=1.  \eea
Since every $g$ has a unique decomposition into elements whose orders
are powers of prime numbers and this decomposition commutes
with forming equivalence classes, this gives us an algorithm
which avoids not only requiring knowledge about the explicit structure
of the chiral ring, but also the use of complex numbers.
In practice it is simpler to substitute $\a_h$ in (\ref{nh}) by the
average of all $\a_h$ in the same class than to replace the summations over
group elements by summations over classes. In our implementation we have used
a mixed strategy.
Of course, as in \cite{aas}, we do not calculate the product over rational
numbers occurring in (\ref{nh}) for each pair $(g,h)$, but only once
for each sets of survivors. To this end we create arrays
whose labels are binary codes for the sets of survivors, where we
intermediately store the contributions from the $\a_h$.
The resulting algorithm is similar to the one described in \cite{aas},
the major difference being the fact that
the occurrence of nontrivial discrete torsions
forces us to go twice over the group, because we have to keep
track of both $g$ and $h$.

The implementation of these concepts in a computer program is straightforward
only in principle:
As in \cite{aas}, one could take the list of 108759 possible skeleton
graphs (for an explanation of this terminology see \cite{nms}) and
generate all phase symmetries. Then one has to generate all possible torsions
consistent with a given symmetry group and at last the corresponding
spectra have to be determined.
In practice calculation time will be enormous unless some care is taken.
We use the following strategy for avoiding redundacy:
In a first step we can eliminate a number of skeletons
for various different reasons.
As we have the complete list of spectra from ADE-type skeletons \cite{ade},
there is no need to redo the calculation of these.
We can also eliminate all invertible skeletons containing D-type
parts because of the A-D-equivalence (this would not work for
non-invertible skeletons because pointers at the D-type part
might spoil the argument).
In addition we can eliminate skeletons with contributions of the type
\beq X_1^2X_2+X_2^2X_1,    \eeq
because any symmetry of such a skeleton would also be a symmetry of
the same skeleton with the above expression replaced by
\beq X_1^3+X_2^3.    \eeq
Whereas the elimination of these cases was done by the computer via a
sorting routine, there were also two skeletons that we eliminated by
hand:
\beq X_1^3+X_2^3+X_3^3+X_4^3+X_5^3+X_6^3+X_7^3+X_8^2X_7+X_9^2X_7   \eeq
is redundant because all of its allowed symmetries can also be realised
by skeletons of the types
\beq X_1^3+X_2^3+X_3^3+X_4^3+X_5^3+X_6^3+X_7^3+X_8^2X_7+X_9^2X_8   \eeq
or
\beq X_1^3+X_2^3+X_3^3+X_4^3+X_5^3+X_6^3+X_7^3+X_8^2X_7+X_9^2X_6,   \eeq
and
\beq X_1^3+X_2^3+X_3^3+X_4^3+X_5^3+X_6^3+X_7^3+X_8^4X_7  \eeq
is redundant because all of its allowed symmetries can also be realised by
\beq X_1^3+X_2^3+X_3^3+X_4^3+X_5^3+X_6^3+X_7^3+X_8^6. \eeq
For the remaining skeletons it is necessary to decide which method for
calculating spectra we want to use.
Whereas the calculation of $\chi$ and $\overline\chi$
without explicit knowledge of the chiral ring is straightforward to
implement, it has the disadvantage of being very slow.
We therefore used a mixed strategy:
For the two skeletons with the largest numbers of models, namely
\beq X_1^3+X_2^3+X_3^3+X_4^3+X_5^3+X_6^3+X_7^2X_9+X_8^2X_7+X_9^2X_8   \eeq
(giving rise to 365120 models) and
\beq X_1^3+X_2^3+X_3^3+X_4^3+X_5^3+X_6^3+X_7^4+X_8^3X_7  \eeq
(183680 models), it is easy to 
find an explicit basis for the chiral
ring, so we used formulas (\ref{charge},\ref{phase}).
In all other cases, which give rise to more than one million models,
we used the formulas for $\c$ and $\5\c$.
Among these other skeletons the one with the greatest number
(namely 37551) of models is
\beq X_1(X_1^2+X_2^2+X_3^2+X_4^2+X_5^2+X_6^2+X_7^2+X_8^2+X_9^2). \eeq
This skeleton has a large permutation symmetry, but it was not necessary to
take advantage of this fact because the group orders and the numbers of 
torsions are not too big (this is a general property of skeletons that
require many additional monomials for non-degeneracy).

\section{Results and conclusions}


\begin{figure} 
\begin{center}{\small	
\begin{tabular}{||rr|r||rr|r||rr|r||rr|r||rr|r||rr|r||}
\hline\hline
$\ng$&$\!\!\na$&$\c$ & $\ng$&$\!\!\na$&$\c$ & $\ng$&$\!\!\na$&$\c$ &
$\ng$&$\!\!\na$&$\c$ & $\ng$&$\!\!\na$&$\c$ & $\ng$&$\!\!\na$&$\c$\\
\hline
48 & 0 & -96	& 26 & 2 & -48	& 13 & 7 & -12	& 8 & 8 & 0	&
	7 & 15 & 16	& 0 & 30 & 60	\\
42 & 0 & -84	& 28 & 4 & -48	& 14 & 8 & -12	& 12 & 12 & 0	&
	3 & 15 & 24	& 5 & 37 & 64	\\
30 & 0 & -60	& 21 & 3 & -36	& 11 & 7 & -8	& 7 & 11 & 8	&
	3 & 19 & 32	& 0 & 42 & 84	\\
31 & 1 & -60	& 15 & 3 & -24	& 13 & 9 & -8	& 9 & 13 & 8	&
	0 & 24 & 48	& 0 & 48 & 96	\\
29 & 1 & -56	& 16 & 8 & -16	& 3 & 3 & 0	& 0 & 6 & 12	&
	4 & 28 & 48	& 1 & 53 & 104	\\
29 & 2 & -54	& 6 & 0 & -12	& 5 & 5 & 0	& 5 & 11 & 12	&
	2 & 29 & 54	& 3 & 57 & 108	\\
24 & 0 & -48	& 11 & 5 & -12	& 6 & 6 & 0	& 7 & 13 & 12	&
	1 & 29 & 56	& 2 & 58 & 112	\\
\hline\hline
\end{tabular} }\\[3mm]
Table I: The 42 ADE spectra with torsion with $p_{01}=p_{10}=0$
\end{center}
\end{figure}

Our calculations result in 3937 different spectra. 2836 of these
spectra come from invertible skeletons and exhibit perfect mirror symmetry,
whereas 846 of the remaining 1101 spectra have no mirror in the present
complete list. Comparing these numbers with the ones for Landau--Ginzburg
orbifolds without torsion \cite{aas}, where we had 3799%
\footnote{
	In \cite{aas} we missed the spectrum
	$(\ng,\na,\c)=(30,10,-40)$, which was first found in \cite{alex},
        because of a 
	programming error. Our current program reproduces that spectrum
        with 
        trivial torsions.}
spectra (2730 of them came
from invertible skeletons and 817 were mirrorless),
we see that we have found only
138 new spectra. These are plotted in figure 1 against the background of
the spectra that can be obtained without torsion.
Among them there are 10 asymmetric spectra with $p_{01}\neq p_{10}$, namely
$\ng=\na\aus\{6,10,12\}$ with $\{p_{01},p_{10}\}=\{0,1\}$,
$\ng=\na=0$ with $\{p_{01},p_{10}\}=\{0,3\}$ and
$\ng=\na=3$ with $\{p_{01},p_{10}\}=\{1,3\}$, which occur already in the ADE
case~\cite{ade}.
The remaining 42 ADE spectra that do not occur for \LGO s without torsion
are listed in table I, whereas
the 86 new non-ADE spectra 
are listed in table II.
They all have $p_{01}=p_{10}=0$.

\begin{figure} 
\begin{center}{\small	
\begin{tabular}{||rr|r||rr|r||rr|r||rr|r||rr|r||rr|r||}
\hline\hline
$\ng$&$\!\!\na$&$\c$ & $\ng$&$\!\!\na$&$\c$ & $\ng$&$\!\!\na$&$\c$ &
$\ng$&$\!\!\na$&$\c$ & $\ng$&$\!\!\na$&$\c$ & $\ng$&$\!\!\na$&$\c$\\
\hline
73 & 4 & -138	&  39 & 5 & -68	 &   33 & 10 & -46 &   17 & 2 & -30
	&   8 & 2 & -12		&   1 & 13 & 24		\\
79 & 10 & -138	&  35 & 2 & -66	 &   22 & 0 & -44  &   19 & 4 & -30
	&   10 & 4 & -12	&   5 & 20 & 30		\\
59 & 2 & -114	&  46 & 13 & -66 &   31 & 9 & -44  &   20 & 5 & -30
	&   9 & 5 & -8		&   7 & 22 & 30		\\
58 & 4 & -108	&  34 & 4 & -60	 &   35 & 13 & -44 &   23 & 8 & -30
	&   12 & 8 & -8		&   8 & 23 & 30		\\
59 & 7 & -104	&  34 & 5 & -58	 &   22 & 1 & -42  &   13 & 1 & -24
	&   14 & 10 & -8	&   30 & 45 & 30	\\
49 & 2 & -94	&  36 & 8 & -56	 &   29 & 8 & -42  &   19 & 8 & -22
	&   13 & 10 & -6	&   6 & 22 & 32		\\
48 & 2 & -92	&  38 & 10 & -56 &   30 & 9 & -42  &   12 & 2 & -20
	&   9 & 7 & -4		&   0 & 18 & 36		\\
45 & 4 & -82	&  44 & 16 & -56 &   22 & 2 & -40  &   16 & 6 & -20
	&   20 & 22 & 4		&   2 & 20 & 36		\\
41 & 1 & -80	&  31 & 4 & -54	 &   18 & 0 & -36  &   21 & 11 & -20
	&   28 & 31 & 6		&   7 & 26 & 38		\\
42 & 2 & -80	&  40 & 13 & -54 &   24 & 6 & -36  &   14 & 5 & -18
	&   5 & 9 & 8		&   11 & 37 & 52	\\
40 & 1 & -78	&  26 & 0 & -52	 &   21 & 4 & -34  &   15 & 6 & -18
	&   2 & 8 & 12		&   5 & 32 & 54		\\
42 & 3 & -78	&  28 & 2 & -52	 &   17 & 1 & -32  &   10 & 2 & -16
	&   13 & 20 & 14	&   13 & 40 & 54	\\
43 & 4 & -78	&  29 & 3 & -52	 &   20 & 4 & -32  &   14 & 6 & -16
	&   5 & 13 & 16		&   4 & 43 & 78		\\
44 & 5 & -78	&  30 & 4 & -52	 &   22 & 6 & -32  &   17 & 9 & -16
	&   9 & 18 & 18		&   4 & 56 & 104	\\
43 & 5 & -76	&  35 & 9 & -52	 &
&&&&&&&&&&&\\
\hline\hline
\end{tabular} }\\[3mm]
Table II: The 86 new spectra that require torsion and do not occur for ADE
models 
\end{center} \end{figure}

Focusing our attention on models with small modulus of the Euler number
$|\c|$, we observe that we found
2 new 2--generation spectra and 2 new 3--generation spectra.
$(13,10,-6)$ is the model with the smallest dimension of the chiral ring among
all 3-generation models that can be obtained as abelian orbifolds of
\LG -models. This model occurs at non--invertible points in the notorious
configuration with 4 fields of weight 1/3 and 3 fields of weight 2/9,
which is also the starting point for Schimmrigk's non-abelian model
with the spectrum $(9,6,-6)$ \cite{rolf}.

Our numbers indicate that the inclusion
of models with non-trivial discrete torsion does not improve mirror symmetry.
This is also reflected in the fact that out of the 817 mirrorless spectra
coming from theories without torsion, only 25 have \naive\ mirrors
(i.e., models with exchanged Hodge numbers) among the models with torsion.
For the models that come from invertible skeletons, on the other hand,
mirror symmetry is perfect 
because the Berglund--H"ubsch
construction \cite{bh} generalizes to the case with torsion \cite{mmi}.

Although considerable care was necessary in the selection of algorithms and
in the organization of the computations, quite a few reserves in improving
efficiency are left. The two major remaining sources of redundancies are
permutation symmetries and
repetitions because of different skeletons giving rise to the
same weights and symmetries.
With some effort 
it should be possible to reduce these effects by a considerable amount.
Hence the complete classification of (2,1) vacua should
be fairly straightforward and even the (2,0) case could be viable.
But from a phenomenological point of view, non-abelian symmetries and
a further breaking of the gauge group are probably more interesting.

{\it Acknowledgements.}  We would like to thank Albrecht Klemm and Alexander
Niemeyer for informing us about their work on abelian orbifolds \cite{alex},
where they first found the spectrum $(30,10,-40)$ that was missing in
\cite{aas}.
The work of M.K. is supported in part by the {\it "Osterreichische
Nationalbank} under grant No. 5026.

\pagebreak

\ni{\Large\bf Figures} \bsk\bsk

\def\figsca{\funit=.7truemm \wi=240 \he=120 \vmul=1}
\def\figlab{\hlab{-60} \hlab{60} \hlab{-120} \hlab{120} \hlab{-180} \hlab{180}
            \vlab{30}  \vlab{60} \vlab{90} }
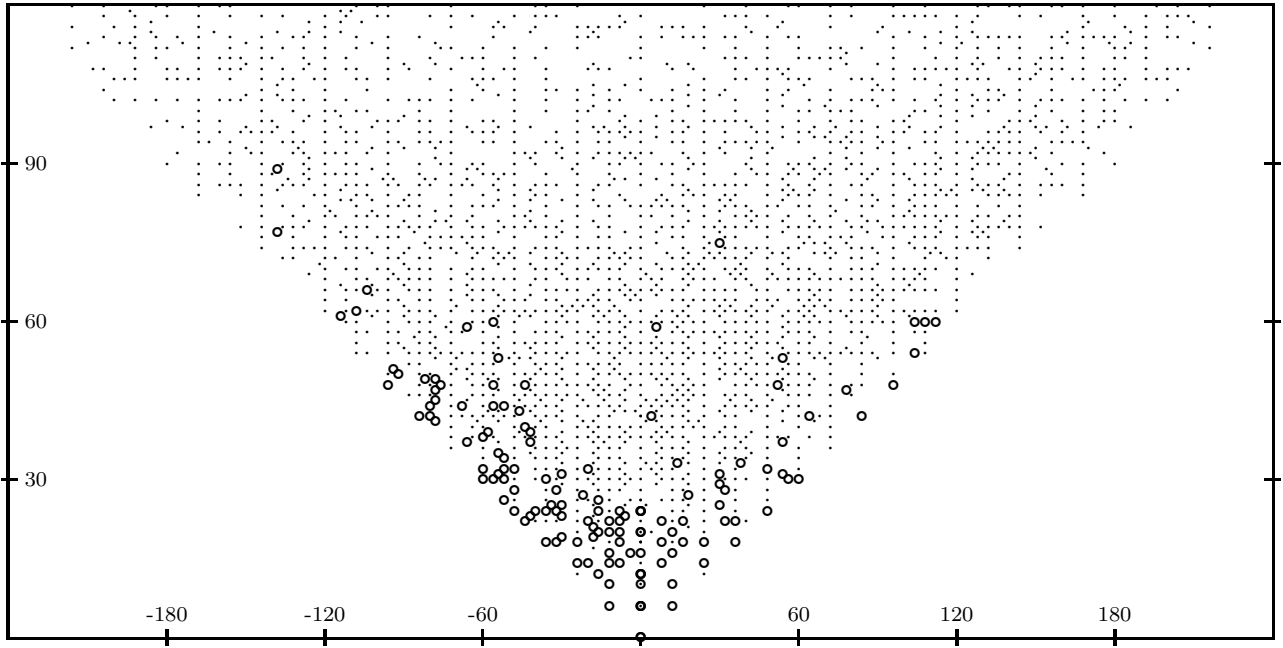
\def\figcap{$\ng+\na$ vs. Euler number for spectra not occurring for any
Landau--Ginzburg\\
orbifold without torsion (circles), and all others with
$\ng+\na\le120$ (dots).}
\Vplo{
\def\npt{{\funit=1pt \circle*1}} 
\.110,2)\.112,4)\.114,6)\.109,4)\.106,2)\.103,1)\.105,3)\.108,6)\.110,8)
\.111,9)\.101,1)\.103,3)\.108,8)\.106,7)\.102,4)\.105,8)\.108,11)\.99,3)
\.101,5)\.102,6)\.103,7)\.104,8)\.105,9)\.106,10)\.107,11)\.95,2)\.104,11)
\.105,12)\.97,5)\.101,9)\.102,10)\.103,11)\.106,14)\.90,0)\.94,4)\.98,8)
\.99,9)\.101,11)\.104,14)\.105,15)\.90,2)\.95,7)\.97,9)\.98,10)\.101,13)
\.92,5)\.100,13)\.96,10)\.84,0)\.85,1)\.86,2)\.87,3)\.89,5)\.90,6)\.91,7)
\.92,8)\.93,9)\.95,11)\.96,12)\.97,13)\.98,14)\.99,15)\.100,16)\.101,17)
\.102,18)\.88,6)\.98,17)\.83,3)\.85,5)\.88,8)\.89,9)\.91,11)\.94,14)\.95,15)
\.100,20)\.82,4)\.83,5)\.84,6)\.85,7)\.86,8)\.90,12)\.92,14)\.93,15)\.94,16)
\.95,17)\.96,18)\.98,20)\.99,21)\.77,1)\.80,4)\.81,5)\.86,10)\.89,13)\.92,16)
\.93,17)\.84,9)\.86,11)\.89,14)\.92,17)\.94,19)\.83,9)\.85,11)\.73,1)\.74,2)
\.75,3)\.76,4)\.77,5)\.78,6)\.79,7)\.80,8)\.81,9)\.82,10)\.83,11)\.84,12)
\.85,13)\.86,14)\.87,15)\.88,16)\.89,17)\.90,18)\.91,19)\.92,20)\.93,21)
\.95,23)\.79,9)\.83,13)\.84,14)\.90,20)\.77,8)\.83,14)\.86,17)\.87,18)\.92,23)
\.70,2)\.73,5)\.75,7)\.77,9)\.79,11)\.81,13)\.85,17)\.87,19)\.88,20)\.90,22)
\.93,25)\.77,10)\.69,3)\.70,4)\.71,5)\.72,6)\.73,7)\.76,10)\.79,13)\.80,14)
\.81,15)\.83,17)\.84,18)\.85,19)\.86,20)\.90,24)\.79,14)\.69,5)\.71,7)\.73,9)
\.75,11)\.76,12)\.77,13)\.78,14)\.79,15)\.81,17)\.83,19)\.86,22)\.87,23)
\.88,24)\.89,25)\.66,3)\.71,8)\.75,12)\.76,13)\.77,14)\.81,18)\.86,23)\.89,26)
\.90,27)\.68,6)\.73,11)\.84,22)\.66,5)\.61,1)\.62,2)\.63,3)\.64,4)\.65,5)
\.66,6)\.67,7)\.68,8)\.69,9)\.70,10)\.71,11)\.72,12)\.73,13)\.74,14)\.75,15)
\.76,16)\.77,17)\.78,18)\.79,19)\.80,20)\.81,21)\.82,22)\.83,23)\.84,24)
\.85,25)\.86,26)\.87,27)\.88,28)\.89,29)\.70,11)\.65,7)\.69,11)\.73,15)
\.75,17)\.79,21)\.62,5)\.67,10)\.70,13)\.71,14)\.77,20)\.78,21)\.86,29)
\.87,30)\.88,31)\.58,2)\.59,3)\.60,4)\.61,5)\.63,7)\.65,9)\.66,10)\.67,11)
\.71,15)\.72,16)\.73,17)\.74,18)\.75,19)\.76,20)\.79,23)\.82,26)\.83,27)
\.87,31)\.69,14)\.87,32)\.54,0)\.55,1)\.56,2)\.57,3)\.60,6)\.61,7)\.62,8)
\.63,9)\.64,10)\.65,11)\.66,12)\.67,13)\.68,14)\.69,15)\.70,16)\.71,17)
\.72,18)\.74,20)\.75,21)\.77,23)\.80,26)\.81,27)\.83,29)\.84,30)\.87,33)
\.85,32)\.53,1)\.55,3)\.57,5)\.60,8)\.63,11)\.65,13)\.66,14)\.68,16)\.69,17)
\.72,20)\.76,24)\.78,26)\.79,27)\.56,5)\.58,7)\.59,8)\.62,11)\.66,15)\.67,16)
\.70,19)\.72,21)\.74,23)\.82,31)\.56,6)\.58,8)\.60,10)\.63,13)\.65,15)\.74,24)
\.84,34)\.85,35)\.68,19)\.49,1)\.51,3)\.52,4)\.53,5)\.54,6)\.55,7)\.56,8)
\.57,9)\.58,10)\.59,11)\.60,12)\.61,13)\.62,14)\.63,15)\.64,16)\.65,17)
\.66,18)\.67,19)\.68,20)\.69,21)\.70,22)\.71,23)\.72,24)\.73,25)\.75,27)
\.77,29)\.78,30)\.79,31)\.80,32)\.81,33)\.83,35)\.84,36)\.58,11)\.67,20)
\.51,5)\.52,6)\.55,9)\.57,11)\.59,13)\.61,15)\.62,16)\.64,18)\.66,20)\.70,24)
\.72,26)\.50,5)\.53,8)\.54,9)\.56,11)\.58,13)\.59,14)\.62,17)\.64,19)\.69,24)
\.72,27)\.77,32)\.45,1)\.48,4)\.49,5)\.52,8)\.53,9)\.55,11)\.56,12)\.57,13)
\.59,15)\.61,17)\.63,19)\.64,20)\.73,29)\.76,32)\.50,7)\.60,17)\.44,2)\.45,3)
\.46,4)\.47,5)\.48,6)\.49,7)\.50,8)\.51,9)\.52,10)\.53,11)\.54,12)\.55,13)
\.56,14)\.57,15)\.60,18)\.61,19)\.62,20)\.63,21)\.64,22)\.65,23)\.66,24)
\.68,26)\.69,27)\.70,28)\.71,29)\.72,30)\.78,36)\.58,17)\.64,23)\.43,3)
\.44,4)\.45,5)\.47,7)\.48,8)\.49,9)\.50,10)\.51,11)\.52,12)\.53,13)\.54,14)
\.55,15)\.56,16)\.57,17)\.58,18)\.59,19)\.63,23)\.67,27)\.68,28)\.69,29)
\.71,31)\.73,33)\.75,35)\.79,39)\.46,7)\.47,8)\.48,9)\.50,11)\.59,20)\.60,21)
\.63,24)\.65,26)\.66,27)\.74,35)\.76,37)\.46,8)\.56,18)\.57,19)\.64,26)
\.66,28)\.75,37)\.48,11)\.58,21)\.36,0)\.37,1)\.38,2)\.39,3)\.40,4)\.41,5)
\.42,6)\.43,7)\.44,8)\.45,9)\.46,10)\.47,11)\.48,12)\.49,13)\.50,14)\.51,15)
\.52,16)\.53,17)\.55,19)\.56,20)\.57,21)\.58,22)\.59,23)\.60,24)\.61,25)
\.62,26)\.63,27)\.64,28)\.65,29)\.66,30)\.67,31)\.68,32)\.69,33)\.70,34)
\.72,36)\.74,38)\.76,40)\.77,41)\.78,42)\.43,8)\.44,9)\.49,14)\.54,19)\.38,4)
\.44,10)\.46,12)\.55,21)\.60,26)\.72,38)\.39,6)\.41,8)\.42,9)\.49,16)\.50,17)
\.51,18)\.54,21)\.56,23)\.60,27)\.61,28)\.65,32)\.35,3)\.36,4)\.37,5)\.39,7)
\.40,8)\.41,9)\.42,10)\.43,11)\.46,14)\.47,15)\.48,16)\.49,17)\.50,18)\.51,19)
\.52,20)\.53,21)\.54,22)\.55,23)\.56,24)\.57,25)\.59,27)\.61,29)\.64,32)
\.67,35)\.69,37)\.71,39)\.75,43)\.36,5)\.52,21)\.60,29)\.68,37)\.32,2)\.33,3)
\.35,5)\.36,6)\.37,7)\.38,8)\.39,9)\.40,10)\.41,11)\.42,12)\.43,13)\.44,14)
\.45,15)\.46,16)\.47,17)\.48,18)\.49,19)\.50,20)\.52,22)\.53,23)\.54,24)
\.56,26)\.57,27)\.59,29)\.60,30)\.62,32)\.63,33)\.64,34)\.67,37)\.68,38)
\.69,39)\.70,40)\.71,41)\.36,7)\.39,10)\.33,5)\.35,7)\.37,9)\.39,11)\.40,12)
\.41,13)\.42,14)\.45,17)\.47,19)\.48,20)\.50,22)\.51,23)\.55,27)\.57,29)
\.58,30)\.61,33)\.62,34)\.63,35)\.67,39)\.71,43)\.72,44)\.32,5)\.34,7)\.35,8)
\.37,10)\.38,11)\.41,14)\.42,15)\.43,16)\.44,17)\.45,18)\.46,19)\.49,22)
\.50,23)\.52,25)\.53,26)\.55,28)\.56,29)\.57,30)\.62,35)\.65,38)\.72,45)
\.34,8)\.44,18)\.45,19)\.47,21)\.49,23)\.52,26)\.57,31)\.67,41)\.71,45)
\.39,14)\.49,24)\.25,1)\.27,3)\.29,5)\.30,6)\.31,7)\.32,8)\.33,9)\.34,10)
\.35,11)\.36,12)\.37,13)\.38,14)\.39,15)\.40,16)\.41,17)\.42,18)\.43,19)
\.44,20)\.45,21)\.46,22)\.47,23)\.48,24)\.49,25)\.50,26)\.51,27)\.52,28)
\.53,29)\.54,30)\.55,31)\.56,32)\.57,33)\.58,34)\.59,35)\.60,36)\.61,37)
\.62,38)\.63,39)\.64,40)\.65,41)\.66,42)\.67,43)\.68,44)\.71,47)\.31,8)
\.34,11)\.38,15)\.42,19)\.30,8)\.33,11)\.34,12)\.36,14)\.39,17)\.42,20)
\.43,21)\.50,28)\.51,29)\.57,35)\.28,7)\.32,11)\.34,13)\.35,14)\.37,16)
\.38,17)\.41,20)\.43,22)\.44,23)\.46,25)\.47,26)\.50,29)\.51,30)\.52,31)
\.53,32)\.56,35)\.59,38)\.68,47)\.21,1)\.25,5)\.26,6)\.27,7)\.28,8)\.29,9)
\.30,10)\.31,11)\.32,12)\.33,13)\.34,14)\.35,15)\.36,16)\.37,17)\.38,18)
\.39,19)\.41,21)\.44,24)\.45,25)\.47,27)\.49,29)\.52,32)\.53,33)\.55,35)
\.56,36)\.65,45)\.68,48)\.69,49)\.70,50)\.30,11)\.32,13)\.35,16)\.20,2)
\.22,4)\.23,5)\.25,7)\.26,8)\.27,9)\.28,10)\.29,11)\.30,12)\.31,13)\.32,14)
\.33,15)\.34,16)\.35,17)\.36,18)\.37,19)\.38,20)\.39,21)\.40,22)\.41,23)
\.42,24)\.43,25)\.44,26)\.45,27)\.46,28)\.47,29)\.48,30)\.49,31)\.50,32)
\.51,33)\.52,34)\.53,35)\.56,38)\.59,41)\.60,42)\.61,43)\.62,44)\.63,45)
\.65,47)\.68,50)\.69,51)\.19,3)\.23,7)\.24,8)\.25,9)\.26,10)\.27,11)\.29,13)
\.30,14)\.31,15)\.32,16)\.33,17)\.34,18)\.35,19)\.36,20)\.37,21)\.38,22)
\.39,23)\.40,24)\.41,25)\.43,27)\.44,28)\.45,29)\.46,30)\.47,31)\.51,35)
\.54,38)\.55,39)\.56,40)\.60,44)\.67,51)\.22,7)\.24,9)\.25,10)\.26,11)\.27,12)
\.28,13)\.29,14)\.30,15)\.32,17)\.33,18)\.34,19)\.36,21)\.38,23)\.39,24)
\.42,27)\.43,28)\.44,29)\.46,31)\.47,32)\.49,34)\.53,38)\.54,39)\.25,11)
\.30,16)\.31,17)\.34,20)\.37,23)\.38,24)\.40,26)\.41,27)\.43,29)\.38,25)
\.51,38)\.12,0)\.16,4)\.17,5)\.18,6)\.19,7)\.20,8)\.21,9)\.22,10)\.23,11)
\.24,12)\.25,13)\.26,14)\.27,15)\.28,16)\.29,17)\.30,18)\.31,19)\.32,20)
\.33,21)\.34,22)\.35,23)\.36,24)\.37,25)\.38,26)\.39,27)\.40,28)\.41,29)
\.42,30)\.43,31)\.44,32)\.45,33)\.46,34)\.47,35)\.48,36)\.49,37)\.50,38)
\.51,39)\.52,40)\.53,41)\.54,42)\.55,43)\.56,44)\.57,45)\.58,46)\.62,50)
\.63,51)\.64,52)\.66,54)\.26,15)\.28,17)\.37,26)\.18,8)\.20,10)\.24,14)
\.25,15)\.27,17)\.29,19)\.31,21)\.33,23)\.34,24)\.37,27)\.39,29)\.40,30)
\.42,32)\.55,45)\.59,49)\.62,52)\.63,53)\.13,4)\.16,7)\.17,8)\.19,10)\.20,11)
\.21,12)\.22,13)\.23,14)\.24,15)\.25,16)\.28,19)\.29,20)\.30,21)\.32,23)
\.33,24)\.35,26)\.36,27)\.37,28)\.40,31)\.44,35)\.46,37)\.47,38)\.50,41)
\.53,44)\.55,46)\.56,47)\.57,48)\.59,50)\.62,53)\.15,7)\.18,10)\.19,11)
\.20,12)\.21,13)\.22,14)\.23,15)\.24,16)\.25,17)\.26,18)\.27,19)\.28,20)
\.29,21)\.30,22)\.31,23)\.32,24)\.33,25)\.34,26)\.35,27)\.36,28)\.37,29)
\.39,31)\.40,32)\.41,33)\.42,34)\.43,35)\.45,37)\.47,39)\.50,42)\.51,43)
\.52,44)\.58,50)\.61,53)\.62,54)\.22,15)\.26,19)\.27,20)\.34,27)\.52,45)
\.12,6)\.15,9)\.16,10)\.17,11)\.18,12)\.19,13)\.20,14)\.21,15)\.22,16)\.23,17)
\.24,18)\.25,19)\.26,20)\.27,21)\.28,22)\.29,23)\.30,24)\.31,25)\.32,26)
\.33,27)\.34,28)\.35,29)\.36,30)\.37,31)\.38,32)\.39,33)\.40,34)\.41,35)
\.42,36)\.43,37)\.44,38)\.46,40)\.47,41)\.48,42)\.49,43)\.50,44)\.51,45)
\.53,47)\.54,48)\.55,49)\.57,51)\.61,55)\.24,19)\.28,23)\.29,24)\.34,29)
\.39,34)\.46,41)\.16,12)\.17,13)\.18,14)\.19,15)\.20,16)\.21,17)\.22,18)
\.24,20)\.25,21)\.27,23)\.28,24)\.29,25)\.30,26)\.31,27)\.32,28)\.33,29)
\.35,31)\.36,32)\.37,33)\.38,34)\.44,40)\.47,43)\.48,44)\.50,46)\.51,47)
\.53,49)\.60,56)\.61,57)\.14,11)\.16,13)\.17,14)\.18,15)\.20,17)\.21,18)
\.23,20)\.24,21)\.25,22)\.26,23)\.29,26)\.30,27)\.31,28)\.32,29)\.34,31)
\.35,32)\.37,34)\.38,35)\.40,37)\.43,40)\.45,42)\.47,44)\.49,46)\.50,47)
\.51,48)\.54,51)\.60,57)\.14,12)\.16,14)\.20,18)\.21,19)\.25,23)\.26,24)
\.28,26)\.29,27)\.31,29)\.33,31)\.34,32)\.37,35)\.38,36)\.42,40)\.47,45)
\.49,47)\.56,54)\.22,21)\.33,32)\.46,45)\.7,7)\.9,9)\.10,10)\.11,11)\.13,13)
\.14,14)\.15,15)\.16,16)\.17,17)\.18,18)\.19,19)\.20,20)\.21,21)\.22,22)
\.23,23)\.24,24)\.25,25)\.26,26)\.27,27)\.28,28)\.29,29)\.30,30)\.31,31)
\.32,32)\.33,33)\.34,34)\.35,35)\.36,36)\.37,37)\.38,38)\.39,39)\.40,40)
\.41,41)\.43,43)\.44,44)\.45,45)\.46,46)\.47,47)\.48,48)\.49,49)\.52,52)
\.53,53)\.55,55)\.56,56)\.57,57)\.59,59)\.12,13)\.16,17)\.20,21)\.22,23)
\.34,35)\.9,11)\.15,17)\.18,20)\.21,23)\.24,26)\.25,27)\.26,28)\.27,29)
\.29,31)\.32,34)\.33,35)\.39,41)\.40,42)\.45,47)\.50,52)\.13,16)\.14,17)
\.15,18)\.16,19)\.17,20)\.18,21)\.19,22)\.20,23)\.21,24)\.23,26)\.26,29)
\.27,30)\.29,32)\.31,34)\.32,35)\.34,37)\.36,39)\.37,40)\.40,43)\.42,45)
\.46,49)\.47,50)\.48,51)\.11,15)\.13,17)\.14,18)\.15,19)\.16,20)\.17,21)
\.18,22)\.19,23)\.20,24)\.21,25)\.22,26)\.23,27)\.24,28)\.25,29)\.26,30)
\.28,32)\.29,33)\.31,35)\.32,36)\.33,37)\.37,41)\.39,43)\.40,44)\.43,47)
\.49,53)\.55,59)\.56,60)\.17,22)\.20,25)\.22,27)\.24,29)\.34,39)\.39,44)
\.52,57)\.6,12)\.8,14)\.9,15)\.10,16)\.11,17)\.12,18)\.13,19)\.14,20)\.15,21)
\.16,22)\.17,23)\.18,24)\.19,25)\.20,26)\.21,27)\.22,28)\.23,29)\.24,30)
\.25,31)\.26,32)\.27,33)\.28,34)\.29,35)\.30,36)\.32,38)\.33,39)\.35,41)
\.36,42)\.37,43)\.38,44)\.40,46)\.41,47)\.42,48)\.44,50)\.45,51)\.46,52)
\.47,53)\.48,54)\.49,55)\.51,57)\.55,61)\.19,26)\.20,27)\.26,33)\.27,34)
\.8,16)\.9,17)\.10,18)\.11,19)\.13,21)\.14,22)\.15,23)\.16,24)\.17,25)\.18,26)
\.19,27)\.20,28)\.21,29)\.22,30)\.23,31)\.24,32)\.25,33)\.26,34)\.27,35)
\.28,36)\.29,37)\.30,38)\.31,39)\.32,40)\.33,41)\.34,42)\.35,43)\.37,45)
\.38,46)\.39,47)\.42,50)\.43,51)\.44,52)\.45,53)\.53,61)\.54,62)\.6,15)
\.8,17)\.10,19)\.11,20)\.12,21)\.13,22)\.14,23)\.16,25)\.17,26)\.18,27)
\.19,28)\.20,29)\.22,31)\.23,32)\.24,33)\.26,35)\.27,36)\.29,38)\.30,39)
\.31,40)\.32,41)\.35,44)\.36,45)\.38,47)\.40,49)\.41,50)\.44,53)\.46,55)
\.47,56)\.48,57)\.50,59)\.15,25)\.18,28)\.19,29)\.22,32)\.24,34)\.25,35)
\.27,37)\.29,39)\.30,40)\.37,47)\.40,50)\.49,59)\.52,62)\.15,26)\.30,41)
\.0,12)\.4,16)\.5,17)\.6,18)\.7,19)\.8,20)\.9,21)\.10,22)\.11,23)\.12,24)
\.13,25)\.14,26)\.15,27)\.16,28)\.17,29)\.18,30)\.19,31)\.20,32)\.21,33)
\.22,34)\.23,35)\.24,36)\.25,37)\.26,38)\.27,39)\.28,40)\.29,41)\.30,42)
\.31,43)\.32,44)\.33,45)\.34,46)\.35,47)\.36,48)\.37,49)\.38,50)\.39,51)
\.40,52)\.41,53)\.42,54)\.43,55)\.44,56)\.45,57)\.46,58)\.47,59)\.50,62)
\.51,63)\.52,64)\.54,66)\.30,43)\.31,44)\.12,26)\.20,34)\.22,36)\.24,38)
\.25,39)\.26,40)\.27,41)\.29,43)\.32,46)\.36,50)\.42,56)\.9,24)\.10,25)
\.12,27)\.14,29)\.17,32)\.18,33)\.19,34)\.20,35)\.21,36)\.23,38)\.24,39)
\.29,44)\.31,46)\.32,47)\.33,48)\.34,49)\.39,54)\.44,59)\.7,23)\.9,25)\.10,26)
\.11,27)\.12,28)\.13,29)\.14,30)\.15,31)\.17,33)\.18,34)\.19,35)\.20,36)
\.21,37)\.22,38)\.23,39)\.24,40)\.25,41)\.27,43)\.28,44)\.29,45)\.30,46)
\.31,47)\.33,49)\.34,50)\.35,51)\.38,54)\.39,55)\.40,56)\.41,57)\.44,60)
\.47,63)\.49,65)\.51,67)\.14,31)\.23,40)\.26,43)\.33,50)\.3,21)\.4,22)\.5,23)
\.6,24)\.7,25)\.8,26)\.9,27)\.10,28)\.11,29)\.12,30)\.13,31)\.14,32)\.15,33)
\.16,34)\.17,35)\.18,36)\.19,37)\.20,38)\.21,39)\.22,40)\.23,41)\.24,42)
\.25,43)\.26,44)\.27,45)\.28,46)\.29,47)\.30,48)\.31,49)\.32,50)\.33,51)
\.34,52)\.35,53)\.38,56)\.40,58)\.41,59)\.42,60)\.44,62)\.45,63)\.46,64)
\.47,65)\.48,66)\.50,68)\.9,28)\.16,35)\.25,44)\.26,45)\.1,21)\.5,25)\.7,27)
\.8,28)\.9,29)\.12,32)\.13,33)\.14,34)\.15,35)\.16,36)\.17,37)\.18,38)\.19,39)
\.20,40)\.21,41)\.24,44)\.25,45)\.26,46)\.27,47)\.29,49)\.30,50)\.31,51)
\.33,53)\.35,55)\.37,57)\.39,59)\.43,63)\.45,65)\.48,68)\.49,69)\.50,70)
\.7,28)\.9,30)\.11,32)\.12,33)\.14,35)\.16,37)\.17,38)\.18,39)\.20,41)\.21,42)
\.23,44)\.26,47)\.29,50)\.30,51)\.34,55)\.35,56)\.36,57)\.38,59)\.47,68)
\.11,33)\.13,35)\.14,36)\.15,37)\.17,39)\.18,40)\.19,41)\.20,42)\.21,43)
\.22,44)\.26,48)\.28,50)\.29,51)\.33,55)\.36,58)\.37,59)\.43,65)\.1,25)
\.2,26)\.3,27)\.5,29)\.6,30)\.7,31)\.8,32)\.9,33)\.10,34)\.11,35)\.12,36)
\.13,37)\.14,38)\.15,39)\.16,40)\.17,41)\.18,42)\.19,43)\.20,44)\.21,45)
\.22,46)\.23,47)\.24,48)\.25,49)\.26,50)\.27,51)\.28,52)\.29,53)\.30,54)
\.31,55)\.32,56)\.33,57)\.34,58)\.35,59)\.36,60)\.37,61)\.38,62)\.39,63)
\.40,64)\.41,65)\.42,66)\.43,67)\.44,68)\.45,69)\.47,71)\.10,35)\.14,39)
\.19,44)\.22,47)\.24,49)\.34,59)\.6,32)\.12,38)\.15,41)\.19,45)\.21,47)
\.28,54)\.41,67)\.45,71)\.7,34)\.8,35)\.10,37)\.11,38)\.12,39)\.14,41)\.15,42)
\.16,43)\.17,44)\.22,49)\.23,50)\.25,52)\.26,53)\.28,55)\.29,56)\.30,57)
\.34,61)\.35,62)\.37,64)\.38,65)\.40,67)\.45,72)\.5,33)\.8,36)\.9,37)\.10,38)
\.11,39)\.12,40)\.13,41)\.15,43)\.16,44)\.17,45)\.19,47)\.20,48)\.21,49)
\.22,50)\.23,51)\.25,53)\.26,54)\.28,56)\.29,57)\.30,58)\.31,59)\.38,66)
\.39,67)\.40,68)\.43,71)\.46,74)\.28,57)\.30,59)\.1,31)\.2,32)\.3,33)\.5,35)
\.6,36)\.7,37)\.8,38)\.9,39)\.10,40)\.11,41)\.12,42)\.13,43)\.14,44)\.15,45)
\.16,46)\.17,47)\.18,48)\.19,49)\.20,50)\.22,52)\.23,53)\.24,54)\.26,56)
\.27,57)\.29,59)\.30,60)\.32,62)\.33,63)\.34,64)\.35,65)\.38,68)\.40,70)
\.41,71)\.42,72)\.45,75)\.15,46)\.16,47)\.22,53)\.37,68)\.3,35)\.7,39)\.8,40)
\.9,41)\.11,43)\.12,44)\.13,45)\.14,46)\.15,47)\.16,48)\.17,49)\.18,50)
\.19,51)\.22,54)\.23,55)\.24,56)\.25,57)\.26,58)\.27,59)\.29,61)\.32,64)
\.34,66)\.35,67)\.37,69)\.39,71)\.4,37)\.8,41)\.10,43)\.12,45)\.14,47)\.15,48)
\.16,49)\.18,51)\.20,53)\.21,54)\.23,56)\.26,59)\.27,60)\.28,61)\.29,62)
\.31,64)\.32,65)\.4,38)\.10,44)\.13,47)\.15,49)\.16,50)\.17,51)\.22,56)
\.27,61)\.38,72)\.9,44)\.14,49)\.19,54)\.20,55)\.27,62)\.30,65)\.33,68)
\.0,36)\.1,37)\.2,38)\.3,39)\.4,40)\.5,41)\.6,42)\.7,43)\.8,44)\.9,45)\.10,46)
\.11,47)\.12,48)\.13,49)\.14,50)\.15,51)\.16,52)\.17,53)\.18,54)\.19,55)
\.20,56)\.21,57)\.22,58)\.23,59)\.24,60)\.25,61)\.26,62)\.27,63)\.28,64)
\.29,65)\.30,66)\.31,67)\.32,68)\.33,69)\.34,70)\.36,72)\.37,73)\.38,74)
\.40,76)\.41,77)\.42,78)\.18,55)\.22,59)\.30,67)\.8,46)\.12,50)\.15,53)
\.19,57)\.28,66)\.31,69)\.36,74)\.37,75)\.8,47)\.11,50)\.18,57)\.20,59)
\.22,61)\.25,64)\.26,65)\.27,66)\.28,67)\.35,74)\.36,75)\.3,43)\.4,44)\.7,47)
\.9,49)\.10,50)\.11,51)\.12,52)\.13,53)\.14,54)\.15,55)\.16,56)\.17,57)
\.18,58)\.19,59)\.23,63)\.25,65)\.27,67)\.28,68)\.29,69)\.30,70)\.31,71)
\.33,73)\.39,79)\.40,80)\.3,45)\.4,46)\.5,47)\.6,48)\.7,49)\.8,50)\.9,51)
\.10,52)\.11,53)\.12,54)\.13,55)\.14,56)\.15,57)\.16,58)\.18,60)\.19,61)
\.20,62)\.21,63)\.22,64)\.23,65)\.24,66)\.25,67)\.26,68)\.27,69)\.28,70)
\.30,72)\.32,74)\.35,77)\.7,50)\.17,60)\.18,61)\.22,65)\.5,49)\.8,52)\.9,53)
\.10,54)\.11,55)\.12,56)\.13,57)\.15,59)\.19,63)\.20,64)\.22,66)\.29,73)
\.30,74)\.31,75)\.4,49)\.8,53)\.9,54)\.10,55)\.11,56)\.13,58)\.14,59)\.17,62)
\.19,64)\.22,67)\.24,69)\.25,70)\.26,71)\.28,73)\.30,75)\.32,77)\.35,80)
\.7,53)\.9,55)\.10,56)\.15,61)\.16,62)\.24,70)\.29,75)\.35,81)\.1,49)\.3,51)
\.4,52)\.5,53)\.6,54)\.7,55)\.8,56)\.9,57)\.10,58)\.11,59)\.12,60)\.13,61)
\.14,62)\.15,63)\.16,64)\.17,65)\.18,66)\.19,67)\.20,68)\.21,69)\.22,70)
\.23,71)\.24,72)\.25,73)\.27,75)\.28,76)\.29,77)\.30,78)\.31,79)\.32,80)
\.33,81)\.35,83)\.36,84)\.6,56)\.8,58)\.9,59)\.10,60)\.14,64)\.15,65)\.19,69)
\.23,73)\.24,74)\.34,84)\.5,56)\.8,59)\.12,63)\.15,66)\.16,67)\.18,69)\.21,72)
\.23,74)\.24,75)\.32,83)\.3,55)\.5,57)\.7,59)\.8,60)\.9,61)\.11,63)\.12,64)
\.13,65)\.14,66)\.15,67)\.16,68)\.17,69)\.19,71)\.24,76)\.25,77)\.27,79)
\.28,80)\.33,85)\.8,61)\.0,54)\.1,55)\.2,56)\.4,58)\.6,60)\.7,61)\.8,62)
\.9,63)\.10,64)\.11,65)\.12,66)\.13,67)\.14,68)\.15,69)\.16,70)\.17,71)
\.18,72)\.20,74)\.21,75)\.23,77)\.26,80)\.27,81)\.29,83)\.30,84)\.32,86)
\.33,87)\.9,64)\.3,59)\.5,61)\.7,63)\.9,65)\.10,66)\.11,67)\.12,68)\.13,69)
\.14,70)\.15,71)\.16,72)\.17,73)\.18,74)\.19,75)\.20,76)\.23,79)\.26,82)
\.27,83)\.31,87)\.8,65)\.10,67)\.11,68)\.14,71)\.16,73)\.17,74)\.20,77)
\.21,78)\.22,79)\.23,80)\.24,81)\.28,85)\.29,86)\.7,65)\.11,69)\.13,71)
\.25,83)\.13,72)\.1,61)\.2,62)\.3,63)\.4,64)\.5,65)\.6,66)\.7,67)\.8,68)
\.9,69)\.10,70)\.11,71)\.12,72)\.13,73)\.14,74)\.15,75)\.16,76)\.17,77)
\.18,78)\.19,79)\.20,80)\.21,81)\.22,82)\.23,83)\.24,84)\.25,85)\.26,86)
\.27,87)\.28,88)\.29,89)\.26,87)\.6,68)\.16,78)\.22,84)\.3,66)\.7,70)\.8,71)
\.11,74)\.12,75)\.13,76)\.14,77)\.18,81)\.19,82)\.20,83)\.23,86)\.26,89)
\.27,90)\.5,69)\.6,70)\.7,71)\.8,72)\.9,73)\.11,75)\.13,77)\.14,78)\.15,79)
\.19,83)\.21,85)\.22,86)\.23,87)\.24,88)\.25,89)\.27,91)\.28,92)\.6,71)
\.10,75)\.14,79)\.22,87)\.3,69)\.4,70)\.5,71)\.6,72)\.7,73)\.9,75)\.10,76)
\.12,78)\.13,79)\.14,80)\.15,81)\.17,83)\.18,84)\.19,85)\.20,86)\.24,90)
\.27,93)\.13,80)\.5,73)\.7,75)\.8,76)\.9,77)\.11,79)\.12,80)\.13,81)\.14,82)
\.17,85)\.18,86)\.21,89)\.22,90)\.23,91)\.5,74)\.8,77)\.9,78)\.14,83)\.17,86)
\.4,74)\.5,75)\.7,77)\.9,79)\.11,81)\.12,82)\.13,83)\.14,84)\.19,89)\.22,92)
\.25,95)\.1,73)\.2,74)\.3,75)\.4,76)\.5,77)\.6,78)\.7,79)\.8,80)\.9,81)
\.10,82)\.11,83)\.12,84)\.13,85)\.14,86)\.15,87)\.16,88)\.17,89)\.18,90)
\.19,91)\.20,92)\.21,93)\.23,95)\.24,96)\.9,83)\.12,86)\.16,90)\.8,83)\.9,84)
\.11,86)\.12,87)\.13,88)\.16,91)\.17,92)\.1,77)\.4,80)\.5,81)\.10,86)\.13,89)
\.16,92)\.17,93)\.21,97)\.6,83)\.10,87)\.4,82)\.5,83)\.7,85)\.8,86)\.11,89)
\.12,90)\.13,91)\.14,92)\.15,93)\.16,94)\.17,95)\.18,96)\.21,99)\.3,83)
\.5,85)\.7,87)\.8,88)\.9,89)\.11,91)\.14,94)\.15,95)\.17,97)\.20,100)\.8,89)
\.16,97)\.17,98)\.6,88)\.12,94)\.0,84)\.1,85)\.2,86)\.3,87)\.5,89)\.6,90)
\.7,91)\.8,92)\.9,93)\.11,95)\.12,96)\.13,97)\.14,98)\.15,99)\.16,100)\.17,101)
\.18,102)\.4,89)\.16,101)\.9,95)\.14,100)\.5,92)\.2,90)\.5,93)\.6,94)\.7,95)
\.8,96)\.9,97)\.10,98)\.13,101)\.12,101)\.0,90)\.4,94)\.5,95)\.7,97)\.8,98)
\.9,99)\.10,100)\.11,101)\.12,102)\.14,104)\.15,105)\.7,98)\.5,97)\.7,99)
\.9,101)\.10,102)\.12,104)\.2,95)\.10,103)\.11,104)\.3,99)\.5,101)\.6,102)
\.7,103)\.8,104)\.9,105)\.10,106)\.11,107)\.4,102)\.11,109)\.1,101)\.3,103)
\.4,104)\.7,107)\.8,108)\.9,109)\.1,103)\.3,105)\.6,108)\.8,110)\.9,111)
\.2,106)\.7,111)\.4,109)\.6,111)\.2,110)\.4,112)\.6,114)\.3,3)\.5,5)\.9,9)
\.13,13)\.21,21)\.9,9) 
\def\npt{\funit=1pt\circle3}  
\.73,4)\.79,10)\.59,2)\.58,4)\.59,7)\.48,0)\.49,2)\.48,2)\.42,0)\.45,4)
\.41,1)\.42,2)\.40,1)\.42,3)\.43,4)\.44,5)\.43,5)\.39,5)\.35,2)\.46,13)
\.30,0)\.31,1)\.34,4)\.34,5)\.29,1)\.36,8)\.38,10)\.44,16)\.29,2)\.31,4)
\.40,13)\.26,0)\.28,2)\.29,3)\.30,4)\.35,9)\.24,0)\.26,2)\.28,4)\.33,10)
\.22,0)\.31,9)\.35,13)\.22,1)\.29,8)\.30,9)\.22,2)\.18,0)\.21,3)\.24,6)
\.21,4)\.17,1)\.20,4)\.22,6)\.17,2)\.19,4)\.20,5)\.23,8)\.13,1)\.15,3)\.19,8)
\.12,2)\.16,6)\.21,11)\.14,5)\.15,6)\.10,2)\.14,6)\.16,8)\.17,9)\.6,0)\.8,2)
\.10,4)\.11,5)\.13,7)\.14,8)\.9,5)\.11,7)\.12,8)\.13,9)\.14,10)\.13,10)
\.9,7)\.3,3)\.5,5)\.6,6)\.8,8)\.12,12)\.20,22)\.28,31)\.5,9)\.7,11)\.9,13)
\.0,6)\.2,8)\.5,11)\.7,13)\.13,20)\.5,13)\.7,15)\.9,18)\.1,13)\.3,15)\.5,20)
\.7,22)\.8,23)\.30,45)\.3,19)\.6,22)\.0,18)\.2,20)\.7,26)\.0,24)\.4,28)
\.11,37)\.2,29)\.5,32)\.13,40)\.1,29)\.0,30)\.5,37)\.4,43)\.0,42)\.0,48)
\.1,53)\.4,56)\.3,57)\.2,58)\.6,6)\.10,10)\.12,12)\.0,0)\.6,6)\.10,10)\.12,12)
\.3,3)\.0,0)\.3,3) 
\def\npt{\funit=1pt\circle*3} 
}


\newpage
\begin{thebibliography}{11}\let\bib=\bibitem
\addtolength{\itemsep}{-3pt}
\bib{sy}A.N.Schellekens and S.Yankielowicz, \npb 330 (1990) 103;\\
    Tables Supplements CERN-TH.5440S/89 and CERN-TH.5440T/89 (unpublished)
\bib{fkss}J.Fuchs, A.Klemm, C.Scheich and M.Schmidt, \plb 232 (1989) 317;\\
    \ap 204 (1990) 1
\bib{z2z2}A.E.Faraggi, \plb326 (1994) 62
\bib{fiqs}A.Font, L.E.Ib\'a\~nez, F.Quevedo and A.Sierra, \npb 337 (1990) 119
\bib{abel}S.A.Abel, C.M.A.Scheich, \plb 312 (1993) 423;
        RAL-94-043 and FTUAM-10/94 preprint, hep-th/9404176
\bib{orbi}L.E.Ib\'a\~nez, J.Mas, H.P.Nilles, F.Quevedo, \npb 301 (1988) 157;\\
	A.Font, L.E.Ib\'a\~nez, F.Quevedo, \plb217 (1989) 272;\\
	T.Mohaupt, MS-TPI-93-09 preprint
\bib{cls}P.Candelas, M.Lynker and R.Schimmrigk, \npb 341 (1990) 383
\bib{lvw}C.Vafa and N.Warner, \plb 218 (1989) 51;\\
    E.Martinec, \plb 217 (1989) 431;\\
    W.Lerche, C.Vafa and N.P.Warner, \npb 324 (1989) 427
\bib{nms}M.Kreuzer and H.Skarke, \npb 388 (1992) 113;\\ \cmp 150, 137 (1992)
\bib{klesch}A.Klemm and R.Schimmrigk, \npb 411 (1994) 559
\bib{aas}M.Kreuzer and H.Skarke, \npb 405 (1993) 305
\bib{v}C.Vafa, \mpl 4 (1989) 1169;
    {\it Superstring Vacua}, preprint HUTP-89/A057
\bib{iv}K.Intriligator and C.Vafa, \npb 339 (1990) 95
\bib{dt}C.Vafa, \npb 273 (1986) 592
\bib{ade}M.Kreuzer and H.Skarke, \plb 318 (1993) 305
\bib{sci}M.Kreuzer and A.N.Schellekens, \npb 411 (1994) 97
\bib{gp}B.R.Greene and M.R.Plesser, \npb 338 (1990) 15
\bib{bh}P.Berglund and T.H"ubsch, \npb 393 (1993) 377;\\
	P.Berglund, M.Henningson, IASSNS-HEP-93-92 preprint, hep-th/9401029
\bib{mmi}M.Kreuzer, \plb 328 (1994) 312
\bib{sum}O.Aharony, S.Yankielowicz and A.N.Schellekens, \npb 418 (1994) 157
\bib{alex}A.Niemeyer, diploma thesis, TU-Munich, August 1993 (unpublished)
\bib{rolf}R.Schimmrigk, \plb 193 (1987) 175
\end{thebibliography}
\end{document}